\documentclass[12pt]{article}

\usepackage{amsmath}

\newlength{\dummysp}
\newcommand{\bbar}[1]{{\overline{#1}}}
\newcommand{\half}{\frac{1}{2}}

\newcommand{\beq}{\begin{eqnarray}}
\newcommand{\eeq}{\end{eqnarray}}
\newcommand{\nnn}{ \nonumber \\ }

\newcommand{\Zbf}{{{\bf Z}}}

\newcommand{\vev}[1]{{\langle #1 \rangle}}

\newcommand{\ord}[1]{{{\cal O}(#1)}}
\newcommand{\gappeq}{\mathrel{\rlap {\raise.5ex\hbox{$>$}}
{\lower.5ex\hbox{$\sim$}}}}
\newcommand{\lappeq}{\mathrel{\rlap{\raise.5ex\hbox{$<$}}
{\lower.5ex\hbox{$\sim$}}}}
\newcommand{\myref}[1]{(\ref{#1})}

\newcommand{\bfe}[1]{\vspace{5pt} {\bf #1 \hspace{1pt}}}

\newcommand{\ben}{\begin{enumerate}}
\newcommand{\een}{\end{enumerate}}

\newcommand{\bit}{\begin{itemize}}
\newcommand{\eit}{\end{itemize}}

\newcommand{\Ncal}{{\cal N}}
\newcommand{\eetee}{E_8 \times E_8}
\newcommand{\tK}{{\tilde K}}

\newcommand{\ti}{\times}
\newcommand{\mo}{(-1)}
\newcommand{\GSM}{$G_{SM}$}
\newcommand{\GNA}{$G_{NA}$}
\newcommand{\tspt}{${\rm spin}(32)/Z_2$}
\newcommand{\so}{$SO(32)$}
\newcommand{\un}[1]{\underline{#1}}

\def\[{\left [}
\def\]{\right ]}
\def\({\left (}
\def\){\right )}

\begin{document}

\begin{titlepage}

\renewcommand{\thefootnote}{\fnsymbol{footnote}}

\hfill Mar.~25, 2004

\hfill hep-ph/0402201

\vspace{0.45in}

\begin{center}
{\bf \Large  Lack of trinification in $Z_3$ orbifolds \\ \vskip 5pt
of the $SO(32)$ heterotic string}
\end{center}

\vspace{0.15in}

\begin{center}
{\bf \large Joel Giedt\footnote{{\tt giedt@physics.utoronto.ca}}}
\end{center}

\vspace{0.15in}

\begin{center}
{\it University of Toronto \\
60 St. George St., \\
Toronto, ON, M5S 1A7 Canada}
\end{center}

\vspace{0.15in}

\begin{abstract}
We report results relating to the trinification scenario
in some explicit string constructions that contain $SU(3)^3$ as
a gauge symmetry.  These models are obtained from symmetric $Z_3$ orbifolds 
of the $SO(32)$ heterotic string with one discrete Wilson line.
We highlight the obstacles that were encountered:
the absence of the usual Higgs sector that would break $SU(3)^3 \to
SU(3)_c \times SU(2)_L \times U(1)_Y$;
the presence of exotics that would generically befoul
gauge coupling unification and lead to fractionally-charged
states in the low energy spectrum.
\end{abstract}

\end{titlepage}

\renewcommand{\thefootnote}{\arabic{footnote}}
\setcounter{footnote}{0}

\bfe{Introduction.}
Here we continue work begun in \cite{Giedt:2003an}.
In that letter, all consistent embeddings with one discrete
Wilson line were enumerated for the case of symmetric
$Z_3$ orbifolds of the $SO(32)$ heterotic string.

In the present work we study the spectra for a few of
the models.  Our main focus will be on models that have $SU(3)^3$ gauge
symmetry.  Given appropriate representations ({\it reprs}), 
this extended gauge symmetry can lead to a {\it trinification}
scenario \cite{Glashow:gc,trinOR}.   

Trinification has been 
suggested as a favorable route
for model-building in explicit string constructions \cite{Kim:2003ch}.
The advantages of this sort of ``unified'' model have
been discussed at length in refs.~\cite{trinOR,Kim:2003ch},
so we only briefly mention them here.  First,
the Higgs representations needed to break the $SU(3)^3$ symmetry
to $G_{SM}=SU(3)\times SU(2) \times U(1)$ are allowed
in affine level 1 constructions, such as the ones
we consider here.  By contrast, the adjoint
Higgses of grand unified theories are not allowed in affine
level 1 constructions \cite{noadj}.  Second, because the electroweak
hypercharge is embedded into $SU(3)^3$, its normalization
is standard, unlike most standard-like 
constructions.\footnote{By ``standard'' normalization,
we mean that the unification of couplings involves
the factor of $\sqrt{5/3}$ in 
$g_3(\Lambda_U) = g_2(\Lambda_U) = \sqrt{5/3} g_Y(\Lambda_U)$,
which is not necessarily the case in heterotic string models
with a standard-like gauge group $G_{SM} \times \cdots$.  See,
for example, Sec.~4 of \cite{Giedt:2001zw} for an elementary discussion
of this generic problem.  Exceptions to the normalization problem
are the standard-like constructions based
on free fermionic models \cite{FF}.}
Third, proton decay can be forbidden
by imposing a $Z_2$ discrete symmetry, while still allowing
for light fermion masses.

We will encounter difficulties realizing the trinification
scenario; the obstruction is due to the absence of the
necessary Higgs reprs to break to \GSM\
{\it along a D-flat direction.}
We point out that a similar result
occurs in the models ($Z_3$ orbifolds of the $\eetee$
heterotic string) constructed in \cite{Kim:2003ch},
though it was overlooked in that case because
D-flatness was not checked.  

In addition to our explorations of the trinification
scenario, we will show that some of the models
enumerated in \cite{Giedt:2003an} are vector-like and
are therefore excluded as extensions of the Standard Model.

\bfe{Accomodating trinification.}
Here we focus on models from \cite{Giedt:2003an} that explicitly 
contain the gauge symmetry $SU(3)^3$, in addition to other factors.  
Using the relations discussed in \cite{Giedt:2003an}, it
is straightforward to show the equivalences
Model 5.6 $\simeq$ Model 2.12 and Model 5.7 $\simeq$ Model 3.13.
The inequivalent models are summarized in Table \ref{hwwe}.

In a trinification scenario, such as has been described
in detail in \cite{trinOR}, we fit the fermion spectrum
of the Standard Model into left-handed fermions 
that fall into the $SU(3)^3$ representation\footnote{Note
that there is an irrelevant change of conventions 
$3 \leftrightarrow \bar 3$ in the second and third $SU(3)$'s
relative to those of \cite{trinOR}.}
\beq
3[(3,\bar 3,1) + (\bar 3,1,3) + (1,3,\bar 3)] 
= 3 \; 27{\rm 's} \; {\rm of}  \; E_6
\label{yier}
\eeq
That this is nothing but the $E_6 \supset SU(3)^3$
decomposition of three $27$'s has been indicated.
Thus \myref{yier}
contains some extra states beyond the Standard Model,
as will be made explicit below.
In what follows, we will use the $E_6$ repr notation where
convenient, although the gauge symmetry is never extended to $E_6$.

For the $\Ncal=1$ supersymmetric constructions we study,
these fermions have scalar partners and fall into chiral
supermultiplets.
We interpret $SU(3)_1 = SU(3)_c$, so that $(3,\bar 3,1)$ and
$(\bar 3,1,3)$ contain quarks.  We decompose 
\beq
SU(3)_2 \supset SU(2)_L \times U(1)_1, \qquad
SU(3)_3 \supset U(1)_2 \times U(1)_3
\label{ksre}
\eeq
Then we give a vacuum expectation value ({\it vev}) 
to a scalar in a $(1,3,\bar 3)$ repr
such that $Y \subset U(1)_1 \times U(1)_2 \times U(1)_3$
survives; i.e., the fields contained in \myref{yier}
will have the usual hypercharges with respect to the
surviving $U(1)$.  

Explicitly, we can decompose the irreducible
reprs of \myref{yier} as follows:
\beq
\Xi_1 &=& (3,\bar 3,1) = \binom{Q}{D} \nnn
\Xi_2 &=& (\bar 3,1,3) = (u^c, d^c, D^c) \nnn
\Xi_3 &=& (1,3,\bar 3) = 
\begin{pmatrix} H_u & H_d & L \cr e^c & \nu^c & N \end{pmatrix}
\label{jlsr}
\eeq
where the row index is the $SU(3)_2$ index (where applicable,
the upper 2 components form an $SU(2)_L$ multiplet) and the column
index is the $SU(3)_3$ index.  The fields $Q,u^c,d^c,L,e^c,H_u,H_d$
are the usual chiral supermultiplets of the MSSM, except that
we have three generations of MSSM Higges.
$D$ are down-like fields and $D^c$ their charge conjugates.  $\nu^c$ are
a charge conjugates of right-handed neutrinos, and $N$ are
singlets.  Vevs are given to the scalar components of $\nu^c$ and $N$
in order to break to $G_{SM}$.

The $SU(3)^3$ breaking vev is given at a very high scale, typically
$\ord{10^{16}}$ GeV.  Since we do not wish to break supersymmetry
at this scale, it is necessary that the vev be along a D-flat 
direction.\footnote{Of course, once D-flatness has been ensured,
one must also check F-flatness.}
For this to be true, a vev must also be given to (i) a
scalar in the $(1,\bar 3,3)$ repr, or (ii) to scalars in
the $(1,\bar 3,1)$ and $(1,1,3)$ reprs.  However, 
possibility (ii) would not preserve $U(1)_Y$, so we discard it.
Since $(1,\bar 3,3)$ is not contained in the spectrum of
chiral supermultiplets \myref{yier}, we must add it to
the spectrum.  Moreover, to have an anomaly free spectrum
we must introduce it as part of a vector pair
\beq
\Xi_3 + \Xi_3^c = (1,3,\bar 3) + (1,\bar 3,3)
\label{yiet}
\eeq
These are the minimal Higgs fields of the supersymmetric
trinification model.  
More such pairs could of course be introduced.  Indeed,
realistic mass spectra for the light fields typically
requires that several $SU(3)^3$ Higgses be introduced \cite{trinOR}.
However, in the string-based context, we would like to solve
the string unification problem by having a uniform running
of the gauge couplings above the scale of $SU(3)^3$ breaking.
Then it is necessary to have 
\beq
27 + \bbar{27} = \sum_{i=1}^3 \[ \Xi_i + \Xi_i^c \]
\eeq
instead of \myref{yiet}.  In the models constructed in \cite{trinOR},
the full spectrum is $8 (27) + 5 (\bbar{27})$.

Given the number of singlets that occur in the string-derived
models considered here, effective Yukawa textures and
hierarchies could presumably be manufactured with
less Higgses than the $5[27 + \bbar{27}]$ employed
in \cite{trinOR}.  Also, exotics in the spectrum will
generically alter the running below the $SU(3)^3$ scale,
so we may have to give up simple unification scenarios
in any case.  Thus we search, minimally, for models that 
have \myref{yier} plus some nonzero
number of vector pairs \myref{yiet}.  For the models of
Table \ref{hwwe} we will find that even this is not possible.

\begin{table}
\begin{center}
\begin{tabular}{cccc}
Model no. & $3V$ & $3a_1$ & \GNA \\
\hline
2.12 & $(1^6,0^{10})$ & $(1^3,0^3,1^3,0^7)$ & $SU(3)^3 \ti SO(14)$ \\
2.21 & $(1^6,0^{10})$ & $(-2,1^2,\mo^3,1^3,0^7)$ & $SU(3)^3 \ti SO(14)$  \\
3.13 & $(1^{12},0^{4})$ & $(1^6,\mo^3,0^3,1^3,0)$  & $SU(3)^3 \ti SU(6)$ \\
3.24 & $(1^{12},0^{4})$ & $(-2,1^2,\mo^3,0^6,1^3,0)$ & $SU(3)^3 \ti SU(6)$ \\
5.23 & $(-2,1^8,0^7)$ & $(-2,1^2,\mo^3,0^3,1^3,0^4)$ & $SU(3)^4 \ti SO(8)$ \\
\hline
\end{tabular}
\caption{Inequivalent models, with an $SU(3)^3$ factor, from \cite{Giedt:2003an}.
Here, $V$ is the twist embedding and $a_1$ is the nonvanishing
Wilson line.  ``Exponents'' indicate the number of times that a
particular entry is repeated.  \GNA\ denotes the nonabelian factors
in the gauge group;
$U(1)$ factors are not shown, but should be supplemented
as required to have rank 16.
\label{hwwe}}
\end{center}
\end{table}

In the models that we study, the untwisted sector has a degeneracy
factor of 3, while the three twisted sectors, denoted by the fixed
point label of the first complex plane, $n_1$, have a degeneracy
factor of 9, since we have only 1 discrete Wilson line.  Thus we
attempt to accomodate trinification by obtaining, say, \myref{yier}
in the untwisted sector, and 9, or 18, etc., copies of \myref{yiet} from
twisted sectors.  In addition, we anticipate exotic reprs such as
(underlining here and below indicates that all possible permutations should
be considered) 
\beq
(\underline{1,1,3}), \quad (\underline{1,1,\bar 3})
\label{jsfe}
\eeq
to appear in some of the models.  Ultimately, it would be important to
find flat directions which remove these exotics at a high
scale while preserving \GSM, since these states lead to
particles with fractional electric charge.  For example, the representation
$(3,1,1)$ is an electrically neutral quark, which
would lead to the observation of hadrons with fractional electric 
charge unless it is supermassive.\footnote{The
mass of an exotic is fixed by the requirement that its relic abundance
be made essentially zero in a typical inflation scenario.
See, for example, \cite{Chang:1996vw} or the end of 
Sec.~5 of \cite{Giedt:2001zw}
for further details.}

\bfe{Spectral analysis.}
The methods for calculating the spectrum in heterotic
orbifolds are well-known \cite{orbOR}.
Here we only review the aspects most pertinent to our
considerations.  See, for example, \cite{Giedt:2002hw} and
refs.~therein for a more detailed discussion.

The representations of the massless spectrum
are characterized in terms of the \tspt\
lattice, which we will denote by $\Lambda$.  This
lattice consists of all 16-vectors of the form
\beq
(n_1,\ldots,n_{16}), \qquad
(n_1 + \half,\ldots,n_{16}+\half),
\label{vfb}
\eeq
subject to the constraints $n_I \in \Zbf$ and
$\sum_I n_I  = 0 \mod 2$.  We remind the reader
that ${\rm spin}(32)$ is the covering group for \so.
The \so\ roots are
\beq
(\un{\pm 1, \pm 1, 0^{14}}).
\label{oiy}
\eeq
Here, signs are not correlated.  
The ``exponent'' indicates that the entry is
repeated 14 times.  Analogous notations will
be used below.

In the untwisted sector we have for massless states
with nontrivial ${\rm spin}(32)/Z_2$ weights:
\beq
K \in \Lambda, \qquad K^2 = 2.
\eeq
The Wilson lines $a_i$ enforce a projection on these
states.  Only those that satisfy
\beq
a_i \cdot K \in \Zbf  \quad \forall \; i=1,3,5
\label{jjer}
\eeq
survive.  Those that do survive fall into three
categories, depending on their inner product with
the twist embedding $V$:
\beq
3 V \cdot K = \left\{
\begin{array}{cc}
0 \mod 3 & \qquad {\rm gauge} \\
1 \mod 3 & \qquad {\rm matter} \\
-1 \mod 3 & \qquad {\rm antimatter}
\end{array}
\right.
\label{hyre}
\eeq
In truth this is a further projection onto states
with differing right-moving quantum numbers.

For the massless twisted states, corresponding to string states
with nontrivial monodromy, we have shifted weights $\tK$ which
satisfy
\beq
\tK^2 = \frac{4}{3} - 2N_L, \qquad
\tK = K + V + \sum_{i=1,3,5} n_i a_i, \qquad
K \in \Lambda.
\label{tms}
\eeq
If left-moving oscillators are excited
in the 6-dimensional compact space, we can
have $N_L = 1/3$ or $2/3$.  
The integers $n_i = 0,\pm 1$ label fixed
point locations in each of the 3 complex planes.
Each twisted state is labeled by a triple
$(n_1,n_3,n_5)$.  Note that $3 \tK \in \Lambda$.

We have applied these formulae to the calculation of the
spectra of the models in Table~\ref{hwwe}.  The results
are presented in Table \ref{hwwf}.
It is obvious from inspection
of this table that none of the models work.  
What is missing are the necessary Higgs
representations \myref{yiet} to break $SU(3)^3 \to G_{SM}$.
{\it Bifundamentals are always strictly chiral.}
It can also be seen that there are many exotics.  Even
if we had the Higgses to break to $G_{SM}$ in the usual
way, many fractionally charged particles would generically occur in the
massless spectrum.

\begin{table}
\begin{center}
\begin{tabular}{cl}
Model no. & \hspace{2in} \GNA\ reprs \\ \hline
2.12 & $3[(3,1,\bar 3,1) + 3(\bar 3, \bar 3,1,1) +3(1,\bar 3,3,1)
+ (1,3,1,14) + 6(3,1,1,1)$ \\ 
& \qquad $ + 7(1,3,1,1) + 3(1,\bar 3,1,1) + 6(1,1,\bar 3,1)
+ 18(1,1,1,1)]$ \\
2.21 & $3[4(\bar 3, \bar 3,1,1) + 4(3,1,\bar 3,1) + 4(1,3,3,1)
+ 3(3,1,1,1) + 3(\bar 3,1,1,1)$ \\
& \qquad $ + 3(1,3,1,1) + 3(1,\bar 3,1,1)
+ 3(1,1,3,1) + 3(1,1,\bar 3,1) + 27(1,1,1,1)]$ \\
3.13 & $3[(3,1,3,1)+(\bar 3,1,1,\bar 6) + (1,1,\bar 3,6) + 6(3,1,1,1)
+ 3(\bar 3,1,1,1)$ \\
& \qquad $+ 3(1,3,1,1) + 3(1,\bar 3,1,1) + 6(1,1,3,1)
+ 3(1,1,\bar 3,1) + 18(1,1,1,1)]$ \\ 
3.24 & $3[(\bar 3,\bar 3,1,1) + (3,1,\bar 3,1) + (1,3,3,1)
+ 3(3,1,1,1) + 3(\bar 3,1,1,1)$ \\
& \qquad $+ 3(1,3,1,1) + 3(1,\bar 3,1,1)
+3(1,1,3,1) + 3(1,1,\bar 3,1) + 2(1,1,1,6)$ \\
& \qquad $+ (1,1,1,\bbar{15}_a) + 27(1,1,1,1)]$ \\
5.23 & $3[(\bar 3,\bar 3,1,1,1) + (3,1,1,\bar 3,1) + (1,3,1,3,1)
+ (1,1,3,1,8_v) + (1,1,3,1,1)$ \\
& \qquad $+ 3(3,1,1,1,1) + 3(\bar 3,1,1,1,1)
+ 3(1,3,1,1,1) + 3(1,\bar 3,1,1,1)$ \\
& \qquad $+ 9(1,1,\bar 3,1,1) 
+ 3(1,1,1,3,1) + 3(1,1,1,\bar 3,1)]$ \\
\hline
\end{tabular} 
\caption{Matter content of the models.
\label{hwwf}}
\end{center}
\end{table}

\bfe{Comparison to $\eetee$ case.}
Indeed, this result is similar to what
occurs in the models constructed in \cite{Kim:2003ch}.
However, there it was overlooked that the set of
vevs suggested---in our notation,
two vevs in the $(1,3,\bar 3)$ repr---is
not a D-flat direction.  For instance, for the $U(1)_1$ that
appears in \myref{ksre}, we would have
\beq
D_{U(1)_1} \propto \vev{ |\tilde \nu_1^c|^2 + |\tilde \nu_2^c|^2
+ |\tilde N_1|^2 + |\tilde N_2|^2}
\label{yuwe}
\eeq
Here the notation indicates the scalar components of the
first and second $(1,3,\bar 3)$ that are supposed
to get vevs [cf.~Eq.~\myref{jlsr}].  Without a conjugate
repr $(1,\bar 3,3)$---which carries opposite
charges with respect to $U(1)_1$ and hence
comes in with opposite sign of the terms in
\myref{yuwe}---it is not possible to cancel this D-term.
Similar arguments hold for the other broken generators.

That a D-flat direction does not exist for vevs
in two $(1,3,\bar 3)$ reprs can be seen another
way, by appealing to the more powerful method of
invariants \cite{Buccella:1982nx}.  
Since an $SU(3)^3$ {\it basic invariant}
cannot be constructed with just two $(1,3,\bar 3)$ reprs,
a D-flat direction does not exist.  As a matter of
fact, a breaking of $SU(3)_2 \to SU(2)$ using just
$3$'s but no $\bar 3$'s corresponds to the so-called
{\it critical orbit.}  This very case has been discussed
previously, and has been shown not to 
be a D-flat direction \cite{Brax:2001an}.

\bfe{Models with $V=0$.}
These are Models 1.1-1.6 from \cite{Giedt:2003an}.  
Model 1.1 is just the $SO(32)$
heterotic string---it has no embedding; the model possesses
no matter.  Models 1.2-1.6 are 9 generation, non-chiral models.
Because $V=0$, there is no untwisted matter, as can
be seen from \myref{hyre}.
In the $n_1=0$ twisted sector there is no gauge-charged
matter because no solutions to \myref{tms} 
exist that have nontrivial 
weight $\tK$.  (It is interesting to note, however,
that there are gauge-neutral double-oscillator, $N_L=2/3$, states in this
sector.)
From \myref{tms} it can be seen that $n_1= \pm 1$ 
are related by a conjugation of the weights,
$\tK \leftrightarrow -\tK$.
Thus the gauge charges in the $n_1=1$ sector are opposite
of those in the $n_1= -1$ sector;  
there is no way to accomodate a chiral
gauge theory such as the MSSM.

\bfe{Conclusions.}
It is encouraging that we have been able to find models
with the reprs \myref{yier}.  However, the Higgs reprs
\myref{yiet} are absent for the class of constructions
considered here.  We have explained how this result is similar to what
occurs in the models constructed in \cite{Kim:2003ch}.

The trinification group $SU(3)^3$ may be
obtained as a decomposition of the gauge group of many
of the other models listed in \cite{Giedt:2003an}.  Thus,
it may be possible to embed the trinification scenario
into a theory with extended gauge symmetry.  However, we
have not yet investigated matter representations of these
other models and whether or not they will provide for
both \myref{yier} and \myref{yiet} simultaneously,
or for the Higgs reprs of the extended gauge symmetry
that would break it to $SU(3)^3$ along a flat direction.

Another alternative is to use some of the extra $U(1)$'s,
that are present in the five models studied here, in the
construction of $Y$.  This generalization of the hypercharge 
embedding might allow for the reprs \myref{jsfe} to be given vevs.
However, the MSSM hypercharge has the feel of a contrivance
in such a scheme; we would also expect to have the usual problems 
with nonstandard hypercharge normalization.

\vspace{15pt}

\noindent
{\bf \Large Acknowledgements}

\vspace{5pt}

\noindent
This work was supported by the National Science and Engineering 
Research Council of Canada and the Ontario 
Premier's Research Excellence Award.


\begin{thebibliography}{9}

\bibitem{Giedt:2003an}
J.~Giedt,
``Z(3) orbifolds of the SO(32) heterotic string: 1 Wilson line  embeddings,''
Nucl.\ Phys.\ B {\bf 671} (2003) 133
[arXiv:hep-th/0301232].


\bibitem{Glashow:gc}
S.~L.~Glashow,
``Trinification Of All Elementary Particle Forces,''
Print-84-0577 (BOSTON),
Published in Providence Grand Unif.1984:0088;
in {\it Proc. Fourth Workshop (1984) on Grand Unification,}
ed. K. Kang et al. (World Scientific, Singapore, 1985), p.~88.


\bibitem{trinOR}
G.~Lazarides, C.~Panagiotakopoulos and Q.~Shafi,
``Supersymmetric Unification Without Proton Decay,''
Phys.\ Lett.\ B {\bf 315} (1993) 325
[Erratum-ibid.\ B {\bf 317} (1993) 661]
[arXiv:hep-ph/9306332];
G.~Lazarides and C.~Panagiotakopoulos,
``MSSM from SUSY trinification,''
Phys.\ Lett.\ B {\bf 336} (1994) 190
[arXiv:hep-ph/9403317];
``Mssm And Large Tan Beta From Susy Trinification,''
Phys.\ Rev.\ D {\bf 51} (1995) 2486
[arXiv:hep-ph/9407286].

\bibitem{Kim:2003ch}
J.~E.~Kim,
``Z(3) orbifold construction of $SU(3)^3$ GUT with $\sin^2 \theta_{W,0} =  3/8$,''
Phys.\ Lett.\ B {\bf 564} (2003) 35
[arXiv:hep-th/0301177].;
K.~S.~Choi and J.~E.~Kim,
``Three family Z(3) orbifold trinification, MSSM and doublet-triplet
splitting problem,''
Phys.\ Lett.\ B {\bf 567} (2003) 87
[arXiv:hep-ph/0305002].

\bibitem{noadj}
A.~Font, L.~E.~Ibanez and F.~Quevedo,
``Higher Level Kac-Moody String Models And Their Phenomenological
Implications,''
Nucl.\ Phys.\ B {\bf 345} (1990) 389;
J.~R.~Ellis, J.~L.~Lopez and D.~V.~Nanopoulos,
``Constraints On Grand Unified Superstring Theories,''
Phys.\ Lett.\ B {\bf 245} (1990) 375.

\bibitem{Giedt:2001zw}
J.~Giedt,
``Spectra in standard-like Z(3) orbifold models,''
Annals Phys.\  {\bf 297} (2002) 67
[arXiv:hep-th/0108244].


\bibitem{FF}
A.~E.~Faraggi, D.~V.~Nanopoulos and K.~j.~Yuan,
``A Standard Like Model In The 4-D Free Fermionic String Formulation,''
Nucl.\ Phys.\ B {\bf 335} (1990) 347;
G.~B.~Cleaver, A.~E.~Faraggi, D.~V.~Nanopoulos and J.~W.~Walker,
``Non-Abelian flat directions in a minimal superstring standard model,''
Mod.\ Phys.\ Lett.\ A {\bf 15} (2000) 1191
[arXiv:hep-ph/0002060];
``Phenomenological study of a minimal superstring standard model,''
Nucl.\ Phys.\ B {\bf 593} (2001) 471
[arXiv:hep-ph/9910230];
``Phenomenology of non-Abelian flat directions in a minimal superstring
standard model,''
Nucl.\ Phys.\ B {\bf 620} (2002) 259
[arXiv:hep-ph/0104091].
G.~B.~Cleaver, A.~E.~Faraggi and D.~V.~Nanopoulos,
``A minimal superstring standard model. I: Flat directions,''
Int.\ J.\ Mod.\ Phys.\ A {\bf 16} (2001) 425
[arXiv:hep-ph/9904301];

\bibitem{Chang:1996vw}
S.~Chang, C.~Coriano and A.~E.~Faraggi,
``Stable superstring relics,''
Nucl.\ Phys.\ B {\bf 477} (1996) 65
[arXiv:hep-ph/9605325].


\bibitem{orbOR}
L.~J.~Dixon, J.~A.~Harvey, C.~Vafa and E.~Witten,
``Strings On Orbifolds,''
Nucl.\ Phys.\ B {\bf 261} (1985) 678;
L.~J.~Dixon, J.~A.~Harvey, C.~Vafa and E.~Witten,
``Strings On Orbifolds. 2,''
Nucl.\ Phys.\ B {\bf 274} (1986) 285;
L.~E.~Ibanez, H.~P.~Nilles and F.~Quevedo,
``Orbifolds And Wilson Lines,''
Phys.\ Lett.\ B {\bf 187} (1987) 25;
L.~E.~Ibanez, J.~Mas, H.~P.~Nilles and F.~Quevedo,
``Heterotic Strings In Symmetric And Asymmetric Orbifold Backgrounds,''
Nucl.\ Phys.\ B {\bf 301} (1988) 157.


\bibitem{Giedt:2002hw}
J.~Giedt,
``Heterotic orbifolds,''
U.C.~Berkeley PhD thesis,
arXiv:hep-ph/0204315.

\bibitem{Buccella:1982nx}
F.~Buccella, J.~P.~Derendinger, S.~Ferrara and C.~A.~Savoy,
``Patterns Of Symmetry Breaking In Supersymmetric Gauge Theories,''
Phys.\ Lett.\ B {\bf 115} (1982) 375;
R.~Gatto and G.~Sartori,
``Consequences Of The Complex Character Of The Internal Symmetry In
Supersymmetric Theories,''
Commun.\ Math.\ Phys.\  {\bf 109} (1987) 327;
C.~Procesi and G.~W.~Schwarz,
``The Geometry Of Orbit Spaces And Gauge Symmetry Breaking In Supersymmetric
Gauge Theories,''
Phys.\ Lett.\ B {\bf 161} (1985) 117;
M.~A.~Luty and W.~I.~Taylor,
``Varieties of vacua in classical supersymmetric gauge theories,''
Phys.\ Rev.\ D {\bf 53} (1996) 3399
[arXiv:hep-th/9506098].

\bibitem{Brax:2001an}
P.~Brax and C.~A.~Savoy,
``Supersymmetric flat directions and analytic gauge invariants,''
arXiv:hep-th/0104077.

\end{thebibliography}
\end{document}